\documentclass{aastex63}

\usepackage{graphicx}
\usepackage{color}
\usepackage{booktabs}
\usepackage{verbatim}
\usepackage{soul}
\usepackage{CJK}
\usepackage{txfonts}

\usepackage[normalem]{ulem} 

\received{-}
\revised{-}
\accepted{July 16, 2020}
\submitjournal{PASP}

\shorttitle{Exocomets from a Solar System Perspective}
\shortauthors{Paul~A.~Str\o m et al.}

\begin{document}
\begin{CJK}{UTF8}{gbsn}

\title{Exocomets from a Solar System Perspective}

\correspondingauthor{Paul~A.~Str\o m (formerly Wilson)}
\email{paul.a.wilson@warwick.ac.uk}

\author[0000-0002-7823-1090]{Paul~A.~Str\o m}
\affiliation{Department of Physics, University of Warwick, Coventry CV4 7AL, UK}

\author[0000-0002-2668-7248]{Dennis~Bodewits}
\affiliation{Department of Physics,  Auburn University, Leach Science Center, Auburn, AL 36832, USA}

\author[0000-0003-2781-6897]{Matthew~M.~Knight}
\affiliation{Department of Physics, United States Naval Academy, 572C Holloway Rd, Annapolis, MD 21402, USA}
\affiliation{Department of Astronomy, University of Maryland, College Park, MD 20742, USA}

\author[0000-0001-9129-4929]{Flavien~Kiefer}
\affiliation{Sorbonne Universit\'{e}, CNRS, UMR 7095, Institut d'Astrophysique de Paris, 98 bis bd Arago, 75014 Paris, France}

\author[0000-0002-5859-1136]{Geraint~H.~Jones}
\affiliation{Mullard Space Science Laboratory, University College London, Holmbury St. Mary, Dorking, Surrey RH5 6NT, UK}
\affiliation{The Centre for Planetary Sciences at UCL/Birkbeck, London, UK}

\author[0000-0001-6527-4684]{Quentin~Kral}
\affiliation{LESIA, Observatoire de Paris, Universit\'{e} PSL, CNRS, Sorbonne Universit\'{e}, Universit\'{e} de Paris, 5 place Jules Janssen, 92195 Meudon, France}

\author[0000-0003-4705-3188]{Luca~Matr\`a}
\affiliation{School of Physics, National University of Ireland Galway, University Road, Galway, Ireland}

\author[0000-0002-4133-5216]{Eva~Bodman}
\affiliation{School of Earth and Space Exploration, Arizona State University, P.O. Box 871404, Tempe, AZ 85287-1404, USA}

\author[0000-0002-9814-9588]{Maria~Teresa~Capria}
\affiliation{Istituto di Astrofisica e Planetologia Spaziali (IAPS), INAF, via del Fosso del Cavaliere, I-00133 Roma, Italy}

\author[0000-0003-2076-8001]{Ilsedore~Cleeves}
\affiliation{University of Virginia, Charlottesville, VA, 22904, USA}

\author[0000-0003-0250-9911]{Alan~Fitzsimmons}
\affiliation{Astrophysics Research Centre, School of Mathematics and Physics, Queen's University Belfast, Belfast, UK}

\author[0000-0002-5234-6375]{Nader~Haghighipour}
\affiliation{Institute for Astronomy, University of Hawaii-Manoa, Honolulu, HI 96825, USA}

\author[0000-0002-9971-4956]{John~H.~D.~Harrison}
\affiliation{Institute of Astronomy, University of Cambridge, Madingley Road, Cambridge CB3 0HA, UK}

\author[0000-0002-0756-9836]{Daniela~Iglesias}
\affiliation{Instituto  de  F\'isica  y  Astronom\'ia,  Facultad  de  Ciencias, and N\'ucleo Milenio de Formaci\'on Planetaria - NPF, Universidad de Valpara\'iso, Av. Gran Breta\~na 1111, 5030 Casilla, Valpara\'iso, Chile}

\author[0000-0003-0065-7267]{Mihkel~Kama}
\affiliation{Tartu Observatory, Observatooriumi 1, T\~{o}ravere 61602, Tartumaa, Estonia}
\affiliation{Department of Physics and Astronomy, University College London, Gower Street, London, WC1E 6BT, UK}

\author[0000-0002-8322-3538]{Harold~Linnartz}
\affiliation{Laboratory for Astrophysics, Leiden Observatory, PO Box 9513, NL2300 RA Leiden, The Netherlands}

\author[0000-0001-7031-8039]{Liton~Majumdar}
\affiliation{School of Earth and Planetary Sciences, National Institute of Science Education and Research, HBNI, Jatni 752050, Odisha, India}

\author[0000-0001-6391-9266]{Ernst~J.~W.~de~Mooij}
\affiliation{Astrophysics Research Centre, School of Mathematics and Physics, Queen's University Belfast, Belfast, UK}

\author[0000-0001-7694-4129]{Stefanie~N.~Milam}
\affiliation{NASA Goddard Space Flight Center, Astrochemistry Laboratory, Code 691, Greenbelt, MD 20771, USA}

\author[0000-0002-9298-7484]{Cyrielle~Opitom}
\affiliation{ESO (European Southern Observatory), Alonso de Cordova 3107, Vitacura, Santiago, Chile}
\affiliation{Institute for Astronomy, University of Edinburgh, Royal Observatory, Edinburgh EH9 3HJ, UK}

\author[0000-0002-4388-6417]{Isabel~Rebollido}
\affiliation{Departamento de F\'isica Te\'orica, Universidad Aut\'onoma de Madrid, Cantoblanco, 28049 Madrid, Spain}

\author[0000-0002-3553-9474]{Laura~K.~Rogers}
\affiliation{Institute of Astronomy, University of Cambridge, Madingley Road, Cambridge CB3 0HA, UK}

\author[0000-0001-9328-2905]{Colin~Snodgrass}
\affiliation{Institute for Astronomy, University of Edinburgh, Royal Observatory, Edinburgh EH9 3HJ, UK}

\author[0000-0002-7853-6871]{Clara~Sousa-Silva}
\affiliation{Earth, Atmospheric, and Planetary Sciences Department, Massachusetts Institute of Technology, Cambridge, MA 02139, USA}

\author[0000-0002-8808-4282]{Siyi~Xu(许\CJKfamily{bsmi}偲\CJKfamily{gbsn}艺)}
\affiliation{ NSF's NOIRLab/Gemini Observatory, 670 N'ohoku Place, Hilo, Hawaii, USA, 96720}

\author[0000-0003-3827-8991]{Zhong-Yi~Lin}
\affiliation{Institute of Astronomy, National Central University, Chung-Li 32054, Taiwan}

\author[0000-0003-0562-6750]{Sebastian~Zieba}
\affiliation{ Max-Planck-Institut f\"ur Astronomie, K\"onigstuhl 17, D-69117 Heidelberg, Germany}
\affiliation{Leiden Observatory, Leiden University, PO Box 9513, 2300 RA Leiden, The Netherlands}



\begin{abstract}
Exocomets are small bodies releasing gas and dust which orbit stars other than the Sun. Their existence was first inferred from the detection of variable absorption features in stellar spectra in the late 1980s using spectroscopy. More recently, they have been detected through photometric transits from space, and through far-IR/mm gas emission within debris disks. As (exo)comets are considered to contain the most pristine material accessible in stellar systems, they hold the potential to give us information about early stage formation and evolution conditions of extra Solar Systems. In the Solar System, comets carry the physical and chemical memory of the protoplanetary disk environment where they formed, providing relevant information on processes in the primordial solar nebula. The aim of this paper is to compare essential compositional properties between Solar System comets and exocomets to allow for the development of new observational methods and techniques. The paper aims to highlight commonalities and to discuss differences which may aid the communication between the involved research communities and perhaps also avoid misconceptions. The compositional properties of Solar System comets and exocomets are summarised before providing an observational comparison between them. Exocomets likely vary in their composition depending on their formation environment like Solar System comets do, and since exocomets are not resolved spatially, they pose a challenge when comparing them to high fidelity observations of Solar System comets. Observations of gas around main sequence stars, spectroscopic observations of "polluted" white dwarf atmospheres and spectroscopic observations of transiting exocomets suggest that exocomets may show compositional similarities with Solar System comets. The recent interstellar visitor 2I/Borisov showed gas, dust and nuclear properties similar to that of Solar System comets. This raises the tantalising prospect that observations of interstellar comets may help bridge the fields of exocomet and Solar System comets.
\end{abstract}



\section{Introduction} \label{sec:intro}
Solar System comets are small bodies containing volatiles which sublimate on close approach to the Sun, creating a cloud of dust and gas. Together with asteroids they are regarded as the unused building blocks of the Solar System and much is to be gained by studying their composition and evolution as they provide important clues to the formation of the Solar System \citep{AHearn2017,Eistrup19}. The accretion of icy solids including comets is a widely accepted formation scenario for the cores of the giant planets \citep[e.g.][]{Lis1993, Pol1996, Lambrechts_2014,Bitsch_2015,Bitsch_2019,Alibert_2018}. At a later stage of planet formation comets may have contributed to the delivery of water likely enriched with complex organic molecules (COMs) to the terrestrial planets in the Solar System, although it is still argued whether or not the comets dominate that later delivery \citep{Har2011,altwegg2015,jin2019,lis2019}.

The comparison between small bodies in the Solar System with those around other stars provides a unique window on planet formation processes that can ultimately help us address questions regarding planet formation, and early complex chemistry. The first unambiguously active interstellar exocomet to pass through the Solar System was recently detected and characterised \citep{Guzik2019, Fitzsimmons2019, Opitom2019b}. The exocomet, known as 2I/Borisov, showed gas, dust and nuclear properties similar to that of Solar System comets, but was enriched in CO \citep{Bodewits2020, Cordiner2020}.

The presence of sublimating small bodies orbiting other stars, commonly referred to as `exocomets', has been inferred ever since variable absorption features were detected in the Ca{\sc ii} lines of the star $\beta$~Pictoris ($\beta$~Pic) by \cite{Ferlet_1987}. The term `FEBs' (Falling Evaporating Bodies) has been used interchangeably with exocomets since \cite{beust1990}. There are four stellar systems with observations showing variable exocomet absorption signatures in several lines or which exhibit a clear photometric signature which is attributed to exocomet activity (see Table\,\ref{tab:exocomets}). There are an additional $\sim30$ systems which show variability in one of the Ca{\sc ii} H or K lines or weak photometric signatures that are suggestive of exocomets (see Table\,\ref{tab:exocomet_candidates}). Recent advances in space based photometers such as the {\it Kepler Space Telescope} and the {\it Transiting Exoplanet Survey Satellite} (TESS) have enabled the photometric detections of exocomets as they transit their host star \citep{Kiefer_2017, rappaport2018, zieba2019, Kennedy_2019} which accurately match predictions on the transit shape and depth \citep{lecavelier1999a,lecavelier1999}. Minute irregular dips in stellar light curves have been interpreted as transits of swarms of exocomets \citep[e.g.][]{wyatt2018}. A large fraction of main sequence stars with cold planetesimal belts exhibit traces of gas \citep[e.g.][]{matra2019a,Kral2017,Moor2017,Marino2016}, interpreted as being released by volatile-rich minor bodies \citep{zuckerman2012,Kral2016}.

The `pollution' of white dwarf (WD) atmospheres has been attributed to the accretion of large extrasolar minor bodies with masses ($10^{16}$--$10^{23}$\,kg) equivalent to Solar System asteroids \citep{farihi2010rocky, girven2012constraints, veras2016post}. Detailed abundance studies \citep[e.g.][]{jura2006carbon,xu2013two, harrison2018polluted} suggest that in the majority of the studied systems the accreting material is volatile-poor and rocky, thus, the material is compositionally similar to Solar System asteroids. However, \citet{Xu2017} found evidence for the accretion of volatile-rich icy material in the WD1425+540 system. The photospheric C, N, and O abundances of WD1425+540 relative to the photospheric refractory abundances (Mg, Si, Fe etc.) were not only consistent with icy material \citep{harrison2018polluted} they also resembled the composition of the Solar System comet 1P/Halley. This suggests that polluted white dwarfs could be useful probes of the bulk composition of exocomets.

Despite the great opportunity for exocometary scientists and the Solar System comet community to learn from each other and exchange experiences and techniques, there has been surprisingly little collaboration between them. This paper is the product of a workshop at the Lorentz Centre in Leiden, the Netherlands, in May 2019, which brought experts from these communities together in the same room to share ideas and exchange knowledge and to foster new insights and develop fresh approaches to a very timely topic. As such, this paper is not meant as a comprehensive review of the different fields. Rather, it should serve as an introduction to some core discoveries, provide access to standard and useful references, identify open issues in both fields, and point out common misconceptions. In Section \ref{sec:properties_comets}, we summarise some key properties of comets in the Solar System. Then in Section \ref{sec:properties_exocomets}, we provide a brief overview of the properties of exocomets. In Section \ref{sec:comparison_comets}, we compare the observational similarities and differences between Solar System comets and exocomets before briefly discussing the discovery of interstellar visitors in Section \ref{sec:interstellar_visitors}. Finally we provide a summary and an outlook in Section \ref{sec:summary_and_outlook}.

\section{Properties of comets in the Solar System}
\label{sec:properties_comets}

\subsection{Ice content and sublimation}
\label{sec:ice_sublimation}

For comets, `activity' usually refers to the sublimation of ices stored in the nucleus. The expanding gas drags small dust (and sometimes ice) particles along. The main ices in most comet nuclei are H$_2$O, CO$_2$, and CO; their relative abundances appear to vary greatly among comets and may be a primordial property \citep{AHearn2012,ootsubo2012,bockelee-morvan2017}. 

These ices have very different sublimation temperatures and can therefore start to effectively sublimate at different distances from the Sun \citep{meech2004,Burke_2010}. The onset of significant effective sublimation of H$_2$O ice from the comet nucleus occurs at 180 K or 2.5 au from the Sun; CO$_2$ at 80 K or out to 13 au; and CO at 25 K or 120 au. It is of note that sublimation does not stop outside these distances but rates drop off exponentially. Consequently, comets can be active at great distances from the Sun \citep{Jewitt_2019}.

Typical water sublimation rates of comets observed from Earth are between 10$^{27}$ - 10$^{30}$ water molecules/s, equivalent to 30 - 30,000 kg/s. Jupiter Family comets typically reach water production rates of order 10$^{28}$ molecules/s; whereas Oort Cloud comets frequently reach 10$^{29}$ molecules/s. Sustained comet water production rates above a level of ${\sim}5 \times 10^{29}$ molecules/s are exceptional \citep{AHearn1995}. The largest reported are for comets C/1975 V1 (West) and C/1995 (Hale-Bopp), which both reached peak water production rates above 10$^{31}$ molecules/s \citep{AHearn1995,biver1997}, though higher production rates likely briefly occur in the case of comets that get extremely close to the Sun. For example, C/2011 W3 (Lovejoy) had production rates ${\sim}10^{31}$ molecules/s despite having a nucleus that was at least an order of magnitude smaller than that of Hale-Bopp \citep{raymond2018}.

The \emph{Rosetta} spacecraft traveled alongside comet 67P/Churyumov-Gerasimenko for two years and found that the percentage of surface that can be considered as "active" varies greatly along the orbit, and is influenced not only by the illumination geometry but also by local structural and compositional properties \citep{Combi2020}. For most comets, only a small fraction of the surface, of order 5\%, contributes to the sublimation activity \citep{AHearn1995}, but there are comets for which significantly more of the surface appears to be active \citep{bodewits14}. Other comets eject large amounts of icy grains that seemingly exaggerate the nucleus's active surface area (such as 103P/Hartley 2;  \citealt{AHearn2011}). 

Comet-like, sublimation-driven activity can occur even if an object contains no water ice. Silicates will start to sublimate at 1000--1500 K, depending on their Mg/Fe content (see Fig. 17 in \citealt{Jones2018}). Such temperatures occur only when objects get within $\sim0.1$\,au of the Sun. Asteroid (3200) Phaethon has displayed faint comet-like activity near its perihelion at 0.14\,au, with a variety of mechanisms suggested for this behaviour \citep{jewitt2010}.

Very little is known about comet interiors and how ices are stored and mixed within them \citep{AHearn2017}. The dust-to-ice ratio of comets is an area of active debate; the \emph{Rosetta} mission to comet 67P/Churyumov-Gerasimenko suggested that its nucleus may contain far less volatile material than was previously assumed for comets, with a refractory to ice ratio of 6 $\pm$ 3 in this object  \citep{fulle2017,fulle2019,Choukroun2020}.  In addition, it is a long-standing question whether comets (and other small bodies such as Centaurs) contain amorphous water ice, whose conversion to crystalline ice could drive activity at large distances from the Sun \citep{meech2004,prialnik1987,Guilbert2012,Agarwal2017}.

Once ejected material has left the vicinity of a Solar System comet's nucleus, it forms the comet's dust and ion tails. Neutral cometary gas is ionised by photoionisation, charge transfer with solar wind protons, and electron impact ionisation \citep{cravens1991}. Solar System comets are immersed in the solar wind, a continuous anti-sunward stream of ionised material flowing at several hundred km per second. Once ionised, this plasma is subsumed by the solar wind, forming an ion tail which can be observed remotely. Ion tails can span several astronomical units in length and often appear blue due to the resonant fluorescent emission of CO$^{+}$ (next section). Cometary dust is accelerated anti-sunward by radiation pressure, generally following the classic dust tail formation process first formulated by \citet{finsonprobstein1968}. There are indications that dust can also be affected by the solar wind, i.e. that it is electrically charged, and is affected by the plasma flowing from the Sun, e.g. \citet{kramer2014}, \citet{price2019}.

\subsection{Chemical and elemental composition of the gas and dust}
\label{sec:composition_dust_gas}

Most of our knowledge about the composition of comets comes from observations of the gas and dust surrounding them. The elemental composition of comet dust has been studied \emph{in situ} by instruments on board spacecraft like \emph{Vega 1 \& 2}, \emph{Giotto}, and \emph{Rosetta} \citep{Cochran2015,Bardyn2017}, in laboratories on Earth by collecting particles in Earth's upper atmosphere \citep{Sykes2004}, and directly from a comet's atmosphere by the Stardust mission \citep{Brownlee2014}. The \emph{Giotto}, \emph{Stardust}, and \emph{Rosetta}-COSIMA results were consistent and showed that approximately 50\% of the mass of the cometary dust is solid organic matter (the other half consisting of minerals) and that the average elemental composition is close to that of the Sun \citep{Bardyn2017} with H and He being notable exceptions (they are more abundant in the Sun). Remote observations of dust are most diagnostic in the mid-infrared, where there are absorption features of minerals, but this has only been possible for relatively few bright comets \citep{Wooden2017}. In optical and near-IR wavelengths, the spectrum of comet dust is mostly featureless and slightly reddened with respect to sunlight. There is a very strong forward scattering phase effect \citep[e.g.][]{marcus07,bertini2017} that can enhance the dust's brightness by a factor 1000 or more at phase angles above 175$^\circ$ \citep{Hui2013}.

Most remote gas-phase chemical abundances are measured with respect to H$_2$O. At distances further than 2.5 au from the Sun, the sublimation rates of water decline quickly. The relative abundance at larger heliocentric distances thus does not represent the ice composition of the nucleus \citep{ootsubo2012}. After H$_2$O, the main components of gas comae are usually CO and CO$_2$ (both 0-30\% with respect to H$_2$O; \citealt{AHearn2012}). Additionally, smaller amounts of other molecules are routinely observed in active comets, including CH$_4$, HCN, CH$_3$OH, H$_2$S, etc. Inventories of molecules are given in dedicated review papers \citep{Altwegg2018,Altwegg_2019,bockelee-morvan2017,Mumma2011}. The ROSINA instrument on board \emph{Rosetta} detected more than 55 different species surrounding 67P/Churyumov-Gerasimenko \citep{Altwegg2018}, including the amino acid glycine \citep{Alt2016}, molecular oxygen \citep{Bieler2015}, and the noble gases Ar and Xe \citep{Bal2015,Marty2017}.

Finally, there are many ions, radicals, and fragment species that are formed by physical reactions in the coma or as products of photodissociation, such as CO$^+$, H$_2$O$^+$, CN, CS, OH, C$_3$, C$_2$ etc. \citep{AHearn1995,Fink2009,Cochran2012}. Many of these species have been observed from the ground in numerous comets because they have relatively large fluorescence efficiencies, are often longer lived than their parents, and they are accessible at multiple wavelengths and with different techniques.  However, the origin of many fragment species is unknown or ambiguous \citep{Feldman2004}. An often encountered question in cometary volatiles is therefore whether they are native to the comet nucleus or formed from other processes on the surface or in the coma.

\subsection{Spectroscopic features of comets}
\label{sec:spectroscopic_features_of_comets}

 Cometary contents have been studied and detected in almost every part of the electromagnetic spectrum (see Table\,\ref{table_of_features} for a summary of features). In this section, we aim to identify the brightest and most useful features, the processes that drive them, and some of the diagnostics they provide.  We recall here that this paper is intended as a broad overview which applies to remote observations of comets, typically at a distance of 1\,au from Earth and the Sun. On a one-by-one case, comets can behave very differently, caused by, for example, different observing circumstances (extremely close approaches to Earth, space missions), or intrinsic physical properties (very low or high production rates, close proximity to the Sun, unusual chemical composition). Given the depth and long history of comet science, this work is also not intended as an exhaustive review, but merely to provide some useful starting points and examples.  There are a number of reviews that provide a more comprehensive overview of the field (including \citealt{Mumma2011,bockelee-morvan2017, Cochran2015, Feldman2004}).

Gases in the coma of Solar System comets emit light throughout the electromagnetic spectrum through different excitation mechanisms. At most wavelengths, the dominant process observed is solar pumped fluorescence where small molecules re-emit upon excitation with high efficiency. A second process is emissive photodissociation, also known as prompt emission, where photodissociation produces excited fragments ([O\,{\sc i}], [N\,{\sc i}], [C\,{\sc i}]; c.f. \citealt{Mckay2013,Opitom2019}). Third, in the inner coma, electron impact dissociation reactions may produce excited atomic and molecular fragments such as H, [O\,{\sc i}], CO, OH, etc. \citep{Feldman2015,Bodewits2016}. Given the right conditions this emission mechanism can reveal small traces of volatiles. In a different setting it also led to the discovery of H$_2$O plumes above the surface of Europa \citep{Roth2014}.

In 1996, \emph{Chandra} and the \emph{Extreme UltraViolet Explorer} (EUVE) unexpectedly discovered that comets can be bright ($>1$\,GW) extreme ultraviolet and X-ray sources \citep{Lisse1996, Krasnopolsky1997}. When highly charged solar wind ions (O$^{8+}$, O$^{7+}$, C$^{6+}$, …) collide with the neutral gas surrounding comets, the ions capture one or more electrons into an excited state. As they cascade to the ground state, they emit X-rays \citep{Cravens1997}. This charge exchange emission has distinct spectral features different from thermally excited plasmas and can be used to help characterise the solar wind plasma and structures such as comets' solar wind bow shocks \citep{Wegmann2004}, the amount of neutral gas present, and possibly even its composition \citep{Bodewits2007,Mullen2017}.

In the far- and mid-ultraviolet (120 - 300\,nm), several bright emission lines from atoms such as H, C, and O can be detected. The fluorescent Ly-$\alpha$ emission of atomic hydrogen is routinely used to determine comet water production with the \emph{Solar Wind Anisotropies/Solar and Heliospheric Observatory} (SWAN/SOHO) survey instrument (cf. \citealt{Combi2011}). Notable other bright emission features are the CO fourth positive system (130 - 190\,nm; \citealt{Lupu2007}), the forbidden Cameron bands of CO (190 - 280\,nm; \citealt{Weaver1994}), and the CO$_2^+$  feature around 289\,nm \citep{Festou1981}.

At near-UV and optical wavelengths (300 - 700\,nm), the most commonly observed lines are from fragment species (OH, CN, C$_2$, etc.), which emit at easily accessible optical wavelengths and have very high fluorescence efficiencies. Several high-quality line atlases are available for optical spectra (including \citealt{Cochran2002}, \citealt{Cremonese2002}, and \citealt{Brown1996}). Comet-specific narrowband filters \citep{Farnham2000, Schleicher2004} are commonly used to image the distribution of fragment species in the coma and to survey production and mixing rates \citep{AHearn1995}.

The strongest feature in this region is the OH A\ $^2\Sigma^+$ - X\ $^2\Pi_i$ (0-0) band around 308.5\,nm but due to significant atmospheric extinction in the near-UV and limited telescope facilities that are sufficiently blue sensitive, it is often challenging to study. Instead, the strong CN B\ $^2\Sigma^+$ - X\ $^2\Sigma^+$ violet band near 388.3\,nm is generally the most accessible bright cometary feature \citep[c.f.][]{Feldman2004} that was already identified 139 years ago \citep{Huggins1881}. Although the chemical source of CN is debated (several comets have more CN than its presumed parent HCN), its emission has been detected at large distances ($>5$\,au) from the Sun, e.g., \citet{Cochran1991,Schleicher1997}. Another feature that can be very bright is the CO$^+$ A\ $^{2}\Pi$-X\ $^{2}\Sigma$ ``comet tail'' emission between 300 - 500 nm (cf. \citealt{Opitom2019}), although the brightness of this feature tends to vary substantially between comets, presumably due to the wide range of involved CO abundances \citep{Dello_2016}. Also small carbon chain radicals have been observed. In the 19$^\mathrm{th}$ century, \cite{Huggins1881} observed a blue band at 405.2 nm in the spectrum of comet Tebbutt, that later was identified as arising from C$_3$ emission \citep{Douglas_1951}.

Emission from neutral Na may also be seen in Solar System cometary spectra, particularly those of near-Sun comets (\citealt{Jones2018} and references therein). The origin of this species in these comets is undetermined. Occasionally, spectra have Na features at two distinct velocities \citep{Cremonese1999,LeBlanc2008}, possibly suggesting multiple sources. The \emph{Rosetta} spacecraft did detect Na in dust grains \citep{Schulz2015}, but a consensus on a dust and/or nuclear ice source in different comets has not yet been reached (e.g. see \citealt{Cremonese1999} and \citealt{Cremonese2002} for reviews). 
\citet{Ellinger2015} have suggested an ice source of Na. Sodium observations are discussed further in section~\ref{sec:compositional_diff}.
Despite being key species in exocomet studies, the Ca$^{+}$ ion, and the neutral Ca species, have been observed in a few Solar System comets, with the bright sungrazer C/1965 (Ikeya-Seki) being the first where Ca\,{\sc i} and Ca\,{\sc ii} were detected remotely in significant abundances \citep{slaughter69}. Their sources are undetermined.

Parent molecules are generally observed at infrared, submillimeter, and millimeter wavelengths \citep{bockelee-morvan2004,Mumma2011,Cordiner2014}. Symmetric species lacking permanent dipoles and therefore pure rotational spectra (CO$_2$, CH$_4$, C$_2$H$_2$, C$_2$H$_6$, C$_2$H$_4$,...) can only be observed through ro-vibrational transitions in the IR upon asymmetric excitations, while other polar complex volatiles excited by solar radiation and/or collision processes, also exhibit pure rotational transitions at radio wavelengths. Measurements at these wavelengths offer insight into the molecular complexity of cometary comae \citep{Biver2015}, distribution of species throughout the comae \citep{Cordiner2014}, as well as other physical mechanisms such as jet activity, rotation, gas temperature profile, distributed sources, etc. \citep[see][]{Drahus2010,Bonev2013,Cordiner2017}. These techniques are entering a new era in sensitivity, resolution, and bandwidth (allowing the coverage of multiple molecules simultaneously) that will not only significantly advance comet studies within the Solar System, but likely also be employed for future studies of exocomets.

As is the case with other astronomical spectroscopic fields, the complete and accurate interpretation of cometary composition is limited by the availability of high-quality spectral data, such as comprehensive line lists and accurate broadening parameters. Currently, complete spectral data only exists for a fraction of the molecular species that could be present in comets \citep{Sousasilva2019}. As such, cometary spectra are vulnerable to misinterpretation, and molecular detections susceptible to misassignments. Table\,\ref{table_of_features} highlights the complexity of molecular composition of comet spectra, with dozens of molecular species already discovered. It is therefore plausible that many more molecules are present in comets that cannot yet be correctly identified due to lack of spectral data; it is crucial that these fundamental molecular spectra are obtained, either through experimental measurements \citep[e.g.][]{Hitran2016} or theoretical calculations \citep[e.g.][]{Exomol, Sousasilva2019, Fortenberry2019}.

\section{Properties of exocomets}
\label{sec:properties_exocomets}
Detections of known or suspected exocomets cover a wide variety of systems and techniques. We summarise the current small body nomenclature and propose a definition for exocomets, by building on the phenomenological approach used for Solar System comets. We then present the various lines of evidence for exocomets and highlight the most notable hosting systems.

\subsection{Small body nomenclature}
\label{sec:nomenclature}
The properties of comets and other small bodies in the Solar System are not strictly defined by the International Astronomical Union (IAU). Resolution 5A for its General Assembly XXVI in 2006 defined planets and dwarf planets. By exclusion, all other bodies, except satellites, orbiting the Sun were collectively defined as Small Solar-System Bodies. This population includes a broad variety of objects, including asteroids, Trojans, moons, Centaurs and comets. These classes are mostly defined by their dynamical properties such as their location in the Solar System and their relation with Jupiter \citep{levison1996}. In practice, there is a strong phenomenological component to the qualification of small bodies. Objects are identified as comets when they show activity consisting of a cloud of gas and dust (coma) and/or tails of ions and/or dust. In that simple view, comets are active due to the sublimation of volatile ices while asteroids are rocky and do not show such activity. This phenomenological approach is not perfect; a small number of asteroids have been observed to demonstrate activity (i.e. mass loss) through a variety of non-volatile sublimation processes, including impacts, rotational spin up, and in some cases repeated comet-like activity driven by sublimation of refractory material \citep{jewitt2015}. Comets can cease to show activity far away from the Sun or as an evolutionary end-state when volatiles near their surface are depleted \citep{jewitt2004}. Finally, different small body populations in the Solar System may be connected as objects evolve dynamically and can thus technically change classification over time - for example objects that are now in the so called `scattered disk' may become Centaurs and eventually Jupiter Family Comets \citep{bernstein2004, Sarid2019}.

The identification of an object as a comet, asteroid, Centaur, or even Kuiper belt object depends on its location in the Solar System architecture and these terms raise expectations about the nature of the small body considered. In this paper, we use the word `exocomet' to describe comets which orbit other stars than the Sun and which exhibit some form of observable activity such as the release of gas or dust, e.g. through a coma or tails of ions or dust. The term `Falling Evaporating Bodies' has been used extensively as a synonym for exocomets \citep[e.g.][]{lagrange-henri1992, beust1994}, though this term is misleading: the activity in such objects is likely driven by sublimation rather than evaporation, and `falling' implies objects seen only prior to periastron, while objects are now also detected after periastron \citep[e.g.][]{kiefer2014} when they could be described as `rising'. The variable gas absorption signatures described in Section~\ref{sec:spectroscopic_observations} clearly indicate the presence of exocometary gas and are thus characterised as bonafide exocomets. The asymmetrical transit signatures observed photometrically thought to be caused by transiting dust tails (as described in Section~\ref{sec:transit observations}). The presence of trace amounts of cold gas within debris disks surrounding nearby main-sequence stars is the product of the release of gas from exocometary ices (as described in Section~\ref{sec:Exocomet_belts}). Observational evidence has opened up the possibility that exocomets may be accreting onto white dwarfs (WD) \citep[e.g.][]{Xu2017} as observed indirectly through the analysis of WD atmospheres (see Section~\ref{sec:white_dwarf_pollution}). This suggests the accreting bodies might in some cases show compositional properties akin to Solar System comets which indicates the bodies might indeed represent their extra solar equivalents, although the accretion is more likely dominated by asteroids.

For cases where no cometary activity signatures are observed, we suggest the more inclusive term `Extrasolar Small Bodies' (ESBs). The properties of ESBs are not well constrained and we thus caution that what an object is identified as might later change as new information becomes available - a situation not uncommon for small bodies in our Solar System.

\subsection{Evidence for exocomets}
\label{sec:Evidence_exocomets}

\subsubsection{Spectroscopic observations of variable absorption lines}
\label{sec:spectroscopic_observations}

\begin{figure}
   \centering
   \includegraphics[width=9cm]{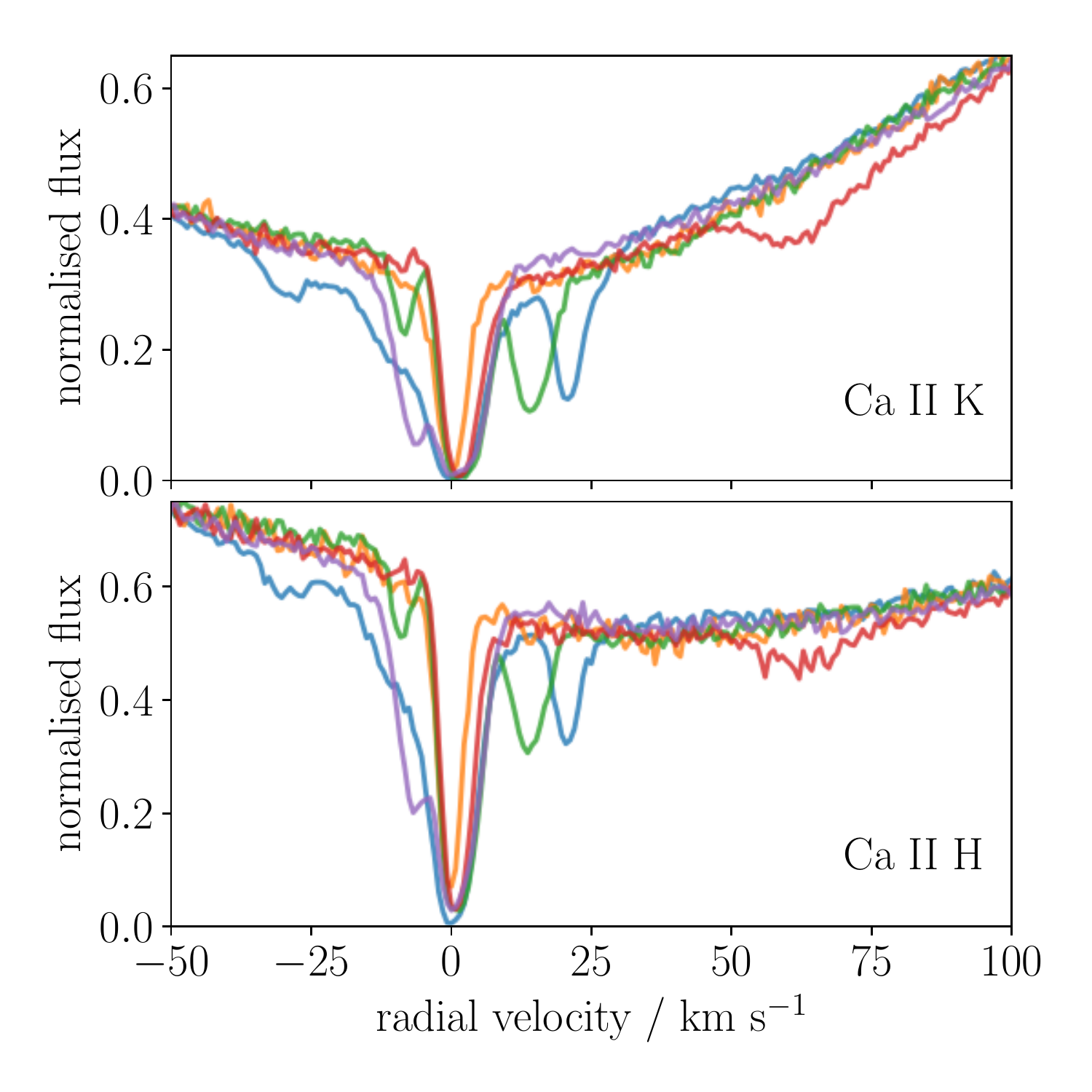}
      \caption{Ca\,{\sc ii} H \& K line of the star $\beta$-Pic. The spectra (each shown in a different colour) were obtained each at a different epoch using data from HARPS.} 
         \label{fig:beta_pic}
   \end{figure}

Exocomets can be discovered spectrocopically by the variable absorption they cause in addition to the stationary stellar Ca\,{\sc ii} H \& K lines (see Fig. \ref{fig:beta_pic}). The absorption occurs when the exocometary gas passes in front of the star, absorbing part of the stellar light as it transits. The interstellar medium also causes absorption in addition to the exocomets. However, this absorption contribution remains constant when compared to the exocomet features which can vary from every few hours to a few times per month. Due to the small equivalent width of these variable exocomet features (few m\AA), only high-resolution instruments (R\,$>$\,60,000; $\Delta v_r<5$\,km\,s$^{-1}$) are suitable for these observations.

$\beta$~Pic is a young ($\sim23$\,Myr) and bright ($V_\mathrm{mag}=4$) A-star which exhibits the largest amount of exocometary activity of any star observed to date. Due to the edge-on inclination of the system as seen from Earth, $\beta$~Pic is well positioned to detect exocomets in the same orbital plane, but much closer to the star than the debris disk observed at tens of au \citep[e.g.][]{Matra2019c}. The orbits of the exocomets around $\beta$~Pic have been estimated indirectly using radial velocity observations combined with a physical sublimation model in an effort to calculate the stellocentric distance at transit \citep{beust1990,Beust1993}. Rapidly varying absorption features (on the order of hours) are observed in sequential radial velocity spectra and are interpreted as being due to accelerating exocomets. The measured acceleration constrains the distance at which the exocomets pass in front of the star which is found to be within a few tens of stellar radii \citep{Kennedy_2018}. These observations are consistent with analytical estimates by \cite{Beust_1996} who estimated that low-velocity Ca\,{\sc ii} absorption features correspond to distances of $\sim15$ to 30 stellar radii ($R_*$) whereas high-velocity features correspond to distances of 5 to 8\,$R_*$. At distances $\lesssim3\ R_*$ the radiation pressure becomes too high for a large and thus detectable cloud to form. This is not the case for other lines such as Mg\,{\sc ii} and Al\,{\sc iii} where the radiation pressure is at least 10 times smaller \citep{Beust_1996}. There is now a large set of $\beta$~Pic spectra, routinely showing deep short term variations in several spectral lines including Ca\,{\sc ii} H \& K in the visible and, for instance, Mg\,{\sc ii} and Fe\,{\sc ii} in the UV. There is a wide variety in the variable features' equivalent width (EW) and velocities. They spread between $-200$ and $+200$\,km/s in radial velocity \citep[see e.g.][]{kiefer2014}, and are in most cases red-shifted compared to the star's doppler shift, i.e. moving towards the central star. The redshift is due to the projection of the velocity of the exocomet onto the line of sight. These properties led the first discoverers of the $\beta$\,Pic phenomenon, to call them "falling evaporating bodies", as discussed earlier. With the exception of the most conspicuous events in $\beta$~Pic, they have EWs of a few m\AA.  No system has been found yet (with the possible exception of $\phi$\,Leo; \citealt{Eiroa16}) with levels of variability comparable to the canonical case of $\beta$~Pic. Their high frequency (hundreds of events per year) implies a large number of objects.

\begin{table}
\begin{center}
\caption{\label{tab:exocomets}Stars with observations showing spectral or photometric variability conclusively attributed to exocomet activity.}   
\begin{tabular}{lcr}
Name       & Sp. Type & Ref.      \\
\hline                    
49 Cet (HD 9672)          & A1V &  (1)       \\
$\beta$~Pic (HD 39060)    & A6V &  (2,3)   \\
HD 172555                 & A7V &  (4)       \\
KIC 3542116 (Photometric detection)  &  F2V &  (5)       \\
\hline
\end{tabular}

\end{center}
References: (1) \cite{Montgomery2012}; (2) \cite{Ferlet_1987}; (3) \cite{kiefer2014}; (4) \cite{Kiefer2014b}; (5) \citep{rappaport2018}.
Spectral types were taken from the references.
\end{table}

\begin{table}
\begin{center}
\caption{\label{tab:exocomet_candidates}Stars which show variability in one of the Ca\,{\sc ii} H or K lines or weak photometric signatures that are suggestive of exocomet activity.}   
\begin{tabular}{lcr}
Name       & Sp. Type & Reference      \\
\hline
HD 256 (HR 10)$^\dagger$  & A2IV/V    &  (1,12,15,20,28) \\
HD 21620                  & A0V       &  (3)       \\
HD 32297                  & A0V       &  (4)       \\
HD 37306 (HR 1919)        & A1V       &  (29)      \\
HD 42111                  & A3V &  (5,12)    \\
HD 50241                  & A7IV      &  (5,11)    \\
HD 56537 ($\lambda$  Gem) & A3V &  (6)       \\
HD 58647                  & B9IV      &  (6)       \\
HD 64145 ($\phi$ Gem)     & A3V &  (6)       \\
HD 80007 (HR 3685)        & A2IV &  (11,15)   \\
HD 85905                  & A2V &  (7,15)    \\
HD 98058 ($\phi$ Leo)     & A5V &  (30)      \\
HD 108767 ($\delta$ Crv)  & A0IV&  (6)       \\
HD 109573 (HR 4796)       & A0V       &  (6,16)    \\
HD 110411  ($\rho$ Vir)   & A0V       &  (3)       \\
HD 138629 (HR 5774)       & A5V       &  (8)       \\
HD 132200 ($\kappa$ Cen)  & B2IV      &  (19)      \\
HD 145964                 & B9V       &  (3)       \\
HD 148283 (HR 6123)       & A5V       &  (5,13)    \\
HD 156623 (HIP 84881)     & A0V       &  (19)      \\
HD 182919 (5 Vul)         & A0V       &  (2)       \\
HD 183324 (c Aql)         & A0IV      &  (10,16)   \\
HD 217782 (2 And)         & A3V &  (2,5,14)  \\
HD 24966                  & A0V       &  (21)      \\
HD 38056                  & B9.5V     &  (21)      \\
HD 79469 ($\theta$ Hya)   & B9.5V     &  (21)      \\
HD 225200                 & A1V       &  (21)      \\
KIC 11084727 (Phot.)      & F2V &  (22)      \\
KIC 8462852 (Phot.)       & F3V       & (23,24,25,26,27) \\
\hline
\end{tabular}

\end{center}
$^\dagger$The star HD 256 (HR 10), while reported in the literature repeatedly as an exocomet-host star, is also a binary with both components hosting a circumstellar stable component \citep[see][]{Montesinos2019}. We present the object in this table to showcase that care must be taken to verify potential false positives and because there still remains unexplained weaker absorptions seen in high resolution spectroscopy by \cite{Welsh1998} which may not correspond to the signatures reported in \cite{Montesinos2019}.\\
References: (1) \cite{Lagrange-Henri1990}; (2) \cite{Montgomery2012};
(3) \cite{Welsh2013}; (4) \cite{Redfield2007a}; (5) \cite{Roberge2008};
(6) \cite{Welsh2015}; (7) \cite{Welsh1998}; (8) \cite{Lagrange-Henri1990b};
(9) \cite{Kiefer2014b}; (10) \cite{Montgomery2017}; (11) \cite{Hempel2003}; (12) \cite{Lecavelier1997};
(13) \cite{Grady1996}; (14)  \cite{Cheng2003}; (15) \cite{Redfield2007b};
(16) \cite{Iglesias18}; (17) \cite{Ferlet_1987};(18) \cite{kiefer2014};
(19) \cite{Rebollido18}; (20) \cite{Eiroa16}; (21) \cite{welsh2018}
(22) \cite{rappaport2018}; (23) \cite{Boyajian16}; (24) \cite{Bodman16}; (25) \cite{Kiefer17}; (26) \cite{Deeg18}; (27) \cite{Wyatt18}; (28) \cite{Montesinos2019}; (29) \cite{Iglesias2019}; (30) \cite{Eiroa16}. Spectral types were taken from the references.
\end{table}

There are three exocomet hosting systems showing variable absorption features attributed to exocomets at both optical and UV wavelengths (see Table\,\ref{tab:exocomets}). The low number of objects is mainly due to the limited number of UV-facilities (recently only the {\it Hubble Space Telescope} (HST) is capable of UV-exocomet observations). Apart from $\beta$~Pic, HD\,172555 \citep{Kiefer2014b,Grady2018} and 49\,Cet \citep{Miles2016}, have shown exocometary-like absorption at both optical and UV wavelengths. Examples include the UV lines Fe~{\sc ii} \citep[e.g.][]{Grady1996,Grady2018} and carbon or oxygen \citep{roberge2006,roberge2014,Grady2018} lines.

A few other systems have showed similar variable spectroscopic features \citep[see e.g.][and references therein]{Eiroa16,Iglesias18,welsh2018} at optical wavelengths. The absorption features have predominantly been found when observing late B and A-type stars. The systems are all young with ages typically ranging from a few tens to hundred of million years old (see \citealt{Wyatt_2007_1,welsh2018} and references therein). There is currently no clear explanation whether this is an observational bias or a physical effect. A possible exception, both in spectral type and age, is the 1.4 Gyr, F2V star $\eta$\,Crv where a tentative absorption signature (2.9 $\sigma$ detection) was detected by \cite{welsh2019}. The majority of targets show possible exocomet induced variability in the Ca\,{\sc ii} H and/or K lines. With a limited number of observational epochs and lower signal to noise, their identification as exocometary, or at least circumstellar in origin is less certain than those of $\beta$~Pic, 49\,Cet and HD\,172555. These candidate exocomet hosting systems which exhibit variability in one of the Ca\,{\sc ii} H or K lines or weak photometric signatures are listed in Table\,\ref{tab:exocomet_candidates}. All stars in this Table require further follow-up to discard periodicity and ensure the variability is not caused by some other process (e.g. stellar pulsations).

\subsubsection{Photometric Transit observations}
\label{sec:transit observations}
\begin{figure}
   \centering
   \includegraphics[width=9cm]{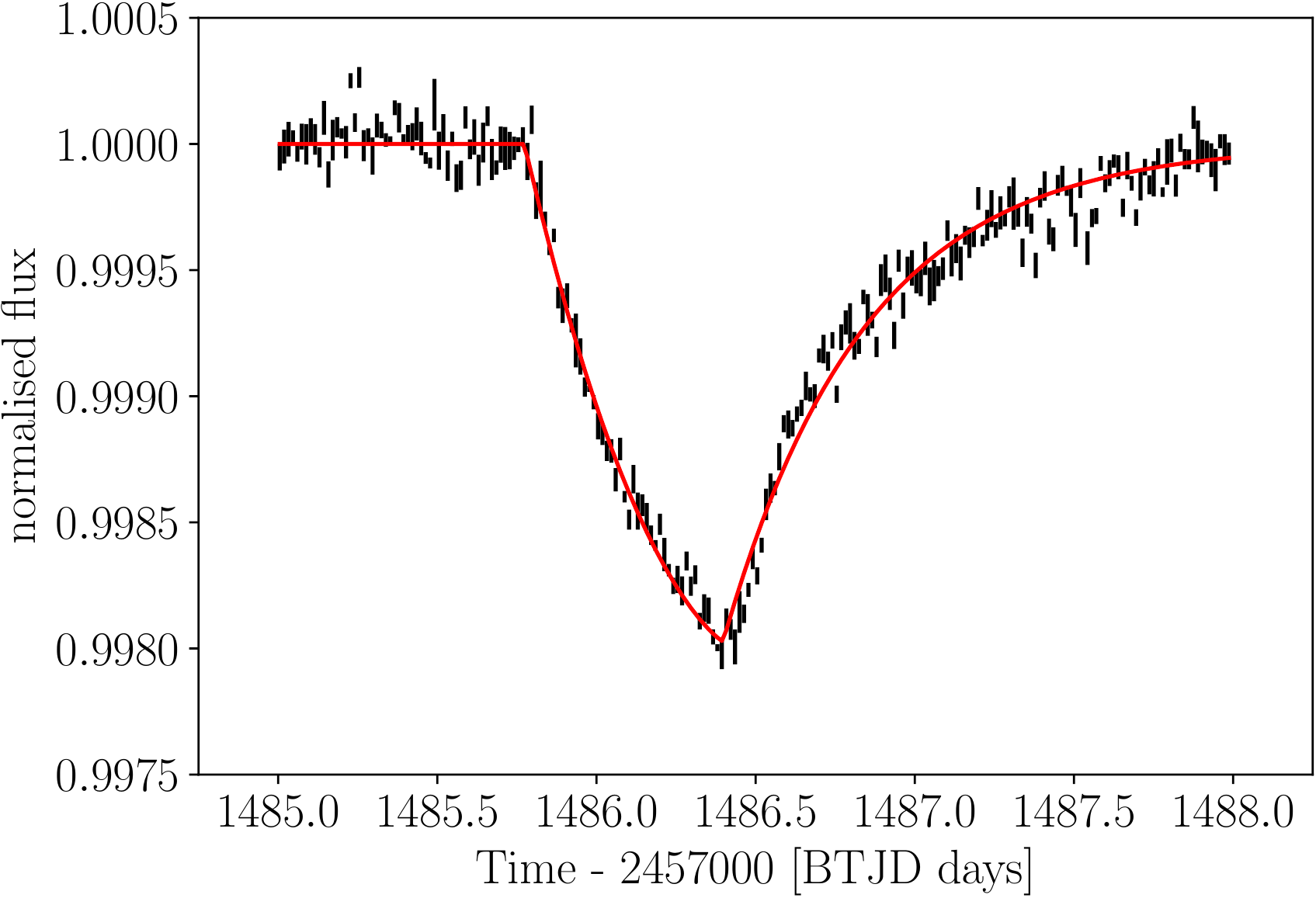}
      \caption{The typical transit shape of an exocomet when observed photometrically. The light curve has been binned to 20 minutes. The red line shows the best-fit model of an exponentially decaying optical depth tail convolved with the limb-darkened disc of the star. Figure adapted from \cite{zieba2019}.}
         \label{fig:exocomets}
   \end{figure}

The milli-magnitude precision of space-based, wide field imagers such as {\it Kepler}/{\it K2} and {\it TESS} have allowed the first detections of exocomets transiting other stars via photometry \citep{rappaport2018,Ansdell2018,zieba2019}. These transit events often have a distinct light curve shape reminiscent of a saw tooth whose shape depends on the angle of the trajectory. The shape was first predicted by \cite{lecavelier1999}, see Fig.\,\ref{fig:exocomets}. The sharp decrease in flux is caused by a steep increase in absorption from the leading edge of the comet's coma followed by an exponential decay back to the full flux level of the star which is caused by the decreasing absorption from the cometary tail. This technique has proven the presence of comet-like bodies around stars with later spectral types (other than A-type), confirming the possible bias in the spectroscopic method. Estimating the size of exocomets is particularly hard due to the large number of degeneracies involved. As long as the exocometary orbital parameters remain unknown there is a large degeneracy between the transverse velocity, the length of the exocomet tail and the impact parameter \citep[see][]{zieba2019}.

\subsubsection{Exocomet populations within debris disks}
\label{sec:Exocomet_belts}
The icy nuclei of exocomets presumably form in the outer regions of protoplanetary disks where conditions are cold enough to allow the freeze-out of volatile molecules and densities are still large enough for the growth of dust up to km-sized comets. These regions are also shielded from radiation from the young stellar object. At the end of planetary assembly, the icy exocomets may be left in one or more belts analogous to the asteroid and Kuiper belts in the Solar System. In the extrasolar context, these reservoirs are known as debris disks, and are commonly observed through the dust and gas which is produced as their members collide and grind down (for a review, see \citealt{Wyatt2008,Hughes_2018}). Debris disks differ substantially from protoplanetary disks. Protoplanetary disks are much younger (typically $< 15$\,Myrs old) and are orders of magnitude more massive in dust and gas, and solid objects are still growing due to the presence of large amounts of primordial gas dominating their dynamics \citep[e.g.][]{wyatt2015}. Debris disks on the other hand exist over a large range of ages and are typically dust-dominated. Observationally, debris disks are generally optically thin, whereas the more dust and gas rich protoplanetary disks are optically thick at visible wavelengths.

The large sensitivity advance and spatial resolution increase brought about by ALMA has recently allowed the detection of cold CO gas within belts around different stellar type stars (19 so far, e.g. \citealt{matra2019a}). Gas has also been detected in atomic form within debris disks, through far-IR/sub-mm emission lines of C\,{\sc i}, C\,{\sc ii} and O\,{\sc i} with Herschel and ALMA \cite[e.g.][]{Cataldi2014,Kral2016,Kral2019}, but also through stable absorption lines seen against UV/optical stellar spectra at the stellar velocity \citep[e.g.][as opposed to red/blue-shifted absorption due to star-grazing exocomets]{Lecavelier_1997}. These are only detectable for belts viewed edge-on \citep[e.g.][]{Rebollido18}, but allow comprehensive compositional inventories, as demonstrated by the large number of volatiles and metallic species detected in gas within the outer $\beta$~Pic belt \citep[e.g.][]{Brandeker2004,roberge2006}.

Multiple component disks are often detected in debris disk systems. Most disks have dust temperatures colder than $\sim150-200$\,K \citep[e.g.][]{Chen_2006,Lawler_2009} although some stars ($\sim 20$\%, see the review by \citealp{Kral2017b}) have a hot dust component ($>300$\,K) within a few au, similar to the asteroid belt or zodiacal dust in the Solar System (see \citealt{Wyatt_2007} and \citealt{Absil_2013}). The F2V star $\eta$ Corvi is a particularly interesting example as it is old ($1.4\pm0.3$ Gyr, \citealt{Nordstrom_2004}) and exhibits both a hot and cold dust component \citep[e.g.][]{Smith2009,Defrere2015}. Using the \emph{Spitzer} space telescope and the Infrared Telescope Facility (IRTF)/SPeX \cite{Lisse_2012} observed $\eta$ Corvi and found that the warm dust spectrum was consistent with very primitive cometary material. They concluded that the parent body or bodies would have been similar to Kuiper Belt Objects in the Solar System which were likely prompted by dynamical stirring to spiral into the inner system. This idea is supported by ALMA observations where CO was detected and thought to originate from volatile rich solid material which sublimates and loses part of its volatiles as it crosses the H$_2$O or CO$_2$ snow lines \citep{Marino_2017}.

While thermal desorption (i.e. sublimation) dominates gas release for Solar System comets approaching the Sun, it is yet unclear which mechanism causes gas release within exocometary belts, since icy Kuiper Belt objects, for example, do not show significant evidence of outgassing \citep{Jewitt2008,Stern2008,Stern2019}. However, extrasolar belts with detected gas are typically much younger (10 to a few 100 Myr-old), and at least 100 times more massive than the current Kuiper belt. They are collision-dominated environments, where km-sized bodies produce a collisional cascade that extends down to micron-sized grains. Then, it is reasonable to assume that gas will also be released within this cascade, for example through resurfacing and the release of trapped volatile material, but also UV-stimulated photodesorption \citep[e.g.][]{Grigorieva2007,Oberg_2009a,Oberg_2009b,Fillion_2014,Martin-Domenech_2015}, or sublimation following high-velocity collisions of accelerated icy grains \citep{Czechowski2007}.

Solar System comets can provide information on the dust properties of debris discs as dust in these disks is considered to be released from exocomets through collisions, i.e. of cometary origin (see \citealt{Hughes_2018}, for a review). Scattered light observations, such as polarimetry, present an excellent opportunity to study the physical properties of the dust particles in orbit around stars as well as cometary dust in the Solar System \citep{Kolokolova2004}. Polarimetric observations yield insights into the distribution of dust grain sizes as well as the spatial distribution from the degree and angle of polarisation as function of wavelength, especially in the case of spatially resolved observations. Polarisation maps of the AU Microscopii (AU Mic) debris disk suggest that the dust particles in the debris disk share a similarly porous structure to cometary dust in the Solar System and that the grains' porosity may be primordial since the dust ring lies beyond the ice sublimation point \citep{Graham_2007}. Analysis of observations of the dust distribution in the $\beta$\,Pic debris disk by \cite{Mirza_2009} showed that a two-disk model fit the data the best and also agree with previously reported disk asymmetries \citep{Heap_2000,Golimowski_2006}. They find that the two disks have dust replenishment times on the order of $\sim10^4$\,yrs at a distance of $\sim100$\,au which hint at the presence of planetesimals that are responsible for the production of second generation dust.
Observations of dust produced through collisions can thus provide a viable way to study exocomets/debris discs whether it be observations of the IR emission (such as the observations of $\eta$ Corvi) or scattered light (such as AU Mic and $\beta$\,Pic).

\subsubsection{White dwarf pollution}
\label{sec:white_dwarf_pollution}
Due to the strong gravitational field, heavy elements are not expected in the atmospheres of WDs \citep{jura2014extrasolar}. Elements heavier than He will sink out of the observable atmosphere on short timescales, much less than the cooling age of the WD \citep{koester2009accretion}. Despite this, between 25--50\% of single WDs are `polluted' by elements heavier than He, which implies the ongoing accretion of material \citep{zuckerman2003metal, zuckerman2010ancient, koester2014frequency}. It has been shown that planetary systems (planets and planetesimal belts) can survive the violent stages of stellar evolution to the WD phase \citep{bonsor2011dynamical, debes2012link, veras2016post}. The standard theory about what causes the `pollution' cites planetesimals being scattered inwards on to star grazing orbits. When these bodies cross the tidal disruption radius they disrupt and subsequently accrete onto the atmosphere of the WD \citep{debes2002there, jura2003tidally, farihi2010rocky, veras2014formation}. 

The mass of the `polluting' bodies has been found to be similar to that of Solar System asteroids ($10^{16}$--$10^{23}$\,kg) by analysing the abundance of metals in the atmospheres of the WDs \citep{xu2012spitzer, girven2012constraints, veras2016post}. However, the exact mass of the `polluting' bodies is difficult to determine as there may be material in a circumstellar reservoir that is yet to be accreted onto the WDs. Furthermore, the accreted material could originate from multiple bodies.

The presence of minor bodies in WD systems is not just implied from spectroscopic studies of the WD photosphere. Comet-like transits of material around WDs can cause large drops in the flux due to the WD's smaller size (compared to main sequence objects), making them easier to identify in the lightcurves. Two WDs have been observed with saw tooth transit features in their lightcurves. WD 1145+017 is a heavily polluted WD with numerous transits. The deepest transit of this WD has a period of roughly 4.5 hours and blocks 60\% of the optical stellar flux \citep{Vanderburg2015}. This is likely from an actively disintegrating minor body in orbit. The WD ZTF J013906.17+524536.89 was found with transits separated by 110 days that caused a 30--45\% drop in the optical stellar flux \citep{vanderbosch2019white}.  Also, one heavily polluted WD (SDSS J122859.93+104033.0) shows evidence for an orbiting planetesimal within its circumstellar gas disc on a $\sim$2 hour orbit \citep{manser2019planetesimal}. 

Spectroscopic observations of the atmospheres of `polluted' WDs can reveal the bulk compositions of the accreted planetary material. To date, 20 different heavy elements have been detected in polluted WDs: C, N, O, Na, Mg, Al, Si, P, S, Ca, Sc, Ti, V, Cr, Mn, Fe, Co, Ni, Cu and Sr \citep{zuckerman2007chemical,xu2013two,melis2016does,Xu2017}. There are more than 20 WDs with detailed abundance analyses and the composition of their pollutants are roughly akin to rocky objects in the Solar System (e.g., \citealt{Gaensicke2012, Klein2010, harrison2018polluted, swan2019interpretation}). So far, there is only one system (WD 1425+540) that has exhibited volatile-rich elements including N, C, and O with an elemental composition similar to the dust surrounding comet 1P/Halley \citep{Xu2017}.

A theoretical study by \cite{veras_2014} found that the delivery of exo-Oort cloud comets onto WDs is dynamically possible, thus, it is not unexpected that white dwarfs can be polluted by such objects. Further evidence for the accretion of comet-like material onto WDs comes from the atmospheres of He dominated WDs (also known as DB WDs). As comet-like material is rich in H, and H unlike all other elements never sinks out of atmosphere of a He dominated WD, the accretion of comet-like material produces a permanent signature in the atmosphere of such a WD. Atmospheric H in He dominated white dwarfs therefore provides evidence for the historical accretion of comet-like bodies. \cite{Hollands_2017,Hollands_2018} analysed the chemical composition of 230 WDs with long cooling ages (T$_\mathrm{eff}$ $<$ 9000K). They found that several of the objects showed large amounts of trace H, thus potentially accreted comet-like material in the past. WD 1425+540 along with the numerous He dominated WDs which contain a significant amount of trace H suggest that WDs may be polluted by analogues of Solar System comets. In our own Solar System the Sun is impacted by comets frequently with comets grazing the Sun every few days \citep{Lamy_2013} compared to asteroids which much less frequently graze the Sun \citep[e.g.][]{Minton_2010,Gladman_1997}. Therefore, it certainly seems possible that exocomets may impact other stars, including WDs. Further studies of polluted WDs may offer a unique insight into the bulk composition of exo-comets.

\subsection{The composition of Exocomets}
\label{sec:Composition_Exocomets}
Gas is released by exocomets transiting their host star at a few stellar radii as well as by the population of exocomets further out at tens of au within cold debris disks, giving us access to their composition. Several gas species attributed to the presence of exocomets have been detected to date (see Table \ref{table_of_exocomet_features}). 

At a few stellar radii, gas is thought to originate from sublimation as exocomets move away from or towards the star, producing red or blue-shifted gas absorption lines. The most readily detected species is Ca$^+$, where absorption is seen in the H\&K lines at 3933.7~\AA\ and 3968.5~\AA, respectively. The absorption signatures typically vary on timescales of hours to days relative to the absorption features caused by circumstellar and interstellar gas which vary on much longer timescales \citep{Kiefer_2019}. Time-variable ultraviolet lines such as Al\,{\sc iii}, C\,{\sc ii}, C\,{\sc iv}, Mg\,{\sc ii}, Fe\,{\sc i} and Fe\,{\sc ii} have been detected in observations of $\beta$\,Pic \citep{Deleuil1993, Vidal-Madjar_1994, Miles2016, Grady2018}, see Table\,\ref{table_of_exocomet_features} for a complete list of species. Although the star itself is unable to photoionise some of these species, it is thought that exocomets sufficiently close to the star (a few stellar radii) form a shock surface where the heat generated by compression and collisions is high enough to produce highly ionised atoms \citep{Beust1993}.

The connection between these red or blue-shifted exocomets at a few stellar radii from their stars and the exocomets that are much more distant orbiting in exo-Kuiper belts (with lines at the stellar velocity, i.e. no shifts) is not clear. The most intuitive interpretation is that we are seeing the same sort of exocomets but at different locations, i.e. the shifted signatures of exocomets appear when the objects are pushed onto small pericentre orbits from large distances. There are dynamical processes that look promising at producing this inward flow of material \citep{beust1996b, beust2000,bonsor2012,bonsor2013,bonsor2014,faramaz2017,Marino2018,sezestre2019}. If this is correct, observations of gas released within debris disks at tens of au can be used to constrain the volatile content, while observations of transits close to the star probe the refractories within exocomets. Observations of exocometary gas within debris disks makes it possible to study the composition of the bulk exocomet population. Assuming the collisional cascade that is producing observable dust is also releasing gas at a steady-state, and that all the ice is lost to the gas phase by the time solids are ground down to the small grains (which are then ejected through radiation pressure), allows derivation of exocometary mass fractions of ice species within the debris disk \citep{zuckerman2012,Matra2015,Kral2017}. In the systems where CO has been measured by ALMA, the CO (and/or CO$_2$) mass fraction in exocomets is consistent with Solar System cometary compositions, within about an order of magnitude \citep{Matra2017}. Searches for gas molecules other than CO are underway with ALMA, though these are harder to detect due to their significantly shorter photodissociation timescales \citep{Matra2018}. Using the excitation of the observed O\,{\sc i} line, \citet{Kral2016} quantified the maximum H$_2$O-to-CO ratio of exocomets within the $\beta$\,Pic debris disk and found that little water is released together with CO; this is consistent with a direct upper limit on H$_2$O gas emission from \textit{Herschel} \citep{cavallius2019}.

Detectable CO release rates observed in debris disks vary by orders of magnitude from $10^{-1}$ to less than $10^{-4}$ M$_\oplus$/Myr. Taking a typical production rate of $10^{-2}$ M$_\oplus$/Myr, we find a production rate of $\sim 10^{34}$ CO molecules/s, which is much higher than what is observed in the Solar System for a given comet (see Sec.\,\ref{sec:ice_sublimation}). This is however not surprising, as the rate in debris disks arises from the collective release from a large amount of exocomets (as first proposed by \citet{Lecavelier_1996} for the $\beta$\,Pic debris disk), rather than the sublimation-driven release by a single object typically observed in the Solar System.

\section{Observational similarities and differences between Solar System comets and Exocomets}
\label{sec:comparison_comets}

\subsection{Spatially resolving (exo)comets}
\label{sec:spatially_resolving_comets}

A major difference between the observations of Solar System comets and exocomets is that the former are studied individually, whereas the latter generally cannot be resolved. Persistent and robotic observing campaigns have accrued chemical abundances for hundreds of comets for fragment species \citep{AHearn1995,Cochran2012, Fink2009}, and for dozens of comets for molecules that are thought to be released directly from the nucleus \citep{AHearn2012,Dello_2016}. These surveys indicate that comets have a broad range of properties that are likely related to their origin in the protoplanetary disk \citep{Davidsson16,Eistrup19}, but that are also  altered by solar heating and interstellar processing. The extent of these effects on comets, and the exact connection between volatiles stored in the nucleus and the observable coma are among the major questions in comet research \citep{AHearn2017, Keller20}. Similar to the comets of the Solar System, exocomet likely also vary in chemical composition depending on their formation environments which is governed by their location in the debris disk and type of star in which they orbit.
 
For exocomets the determination of the properties of individual objects is a challenge as we can never be sure we are only observing a single object or collection of fragments originating from a single object. Some photometric and spectroscopic transits of exocomets indeed suggest the multiplicity of transiting objects \citep{Beust_1996,Neslussan2017,Kiefer_2017}. The exception to this are interstellar visitors (see Sect.\,\ref{sec:interstellar_visitors}) which can be studied individually. Photometric observations which show transit light curves consistent with what is expected from a single exocomet transit may increase the likelihood that a single exocomet is being characterised, although there is currently no method in place for verifying the single nature of photometrically detected exocomets. Observations of the gas in planetesimal belts likely measure the combined composition of large numbers of exocomets. Similarly, the elemental composition derived from WD pollution and spectroscopic observations is likely the product of multiple exocomets and therefore reflects the chemical properties of an entire population of objects.

\subsection{Detection methods}
\label{sec:detection_methods}

Exocomets can be detected through the gas and dust they release, producing absorption of starlight and/or emission. First, exocomets at a few stellar radii can be detected with high resolution spectrographs through their gas coma, which covers a significant fraction of the stellar surface and hence makes them detectable in absorption against the star \citep{lagrange-henri1992}. High resolving power is needed, as delivered by the current generation of optical and NIR echelle spectrographs have spectral resolutions of $120,000$ \citep{Mayor2003} to $190,000$ \citep{Pepe2017}. This allows measurement of the amount of absorption as a function of radial velocity at resolutions of $\sim2$-5\,km/s. This is more than suitable for detecting exocomets which typically display red and blueshifted absorption signatures with a radial velocity in the range from 0 to $\pm200$\,km/s.

Second, with the introduction of sensitive space based photometers such as {\it Kepler}/{\it K2} and {\it TESS}, exocomets close to the star can also be detected by the light blocked out as the dust released from their surface transits the host star. Similar to exoplanets, which display a unique light curve as they transit, transiting exocomets produce their own unique light curve as shown in Section \ref{sec:transit observations}. This provides information about the dust density in the tail. They could also be potentially a source of false positives for single transiting exoplanets when their trajectory causes a symmetric lightcurve \citep[see][]{lecavelier1999}.

Third, exocomets further away from their host stars, orbiting within debris belts, can also be observed through the gas they release as part of the collisional cascade. This gas can be seen in absorption in the UV/optical for edge-on systems \citep[e.g.][]{Brandeker2004,roberge2006} and in emission in the far-IR (e.g. C{\sc ii}, O{\sc i}) or in the sub-mm (e.g. CO, C~{\sc i}).

Solar System comet nuclei are obscured when active, but their nuclei can be directly imaged (optical/IR/radar) when they are far from the Sun, weakly active, and/or very close to Earth. Similar observations of exocomet nuclei will not be possible for the foreseeable future. Even detections of transits by solid bodies will only be possible for bodies much larger than any minor bodies in the Solar System. Currently, the detection of exocomet transits requires much higher optical depth than is seen for typical comets in the Solar System. This suggests either large exocomets ($\sim10-100$ km), larger than most Solar System comets \citep{Bauer_2017} but comparable to many Kuiper Belt objects \citep{Schlichting2013} with extended dust tails transiting the star and/or a system containing a group of exocomets, possibly disintegrating.

\begin{figure}
   \centering
   \includegraphics[width=9cm]{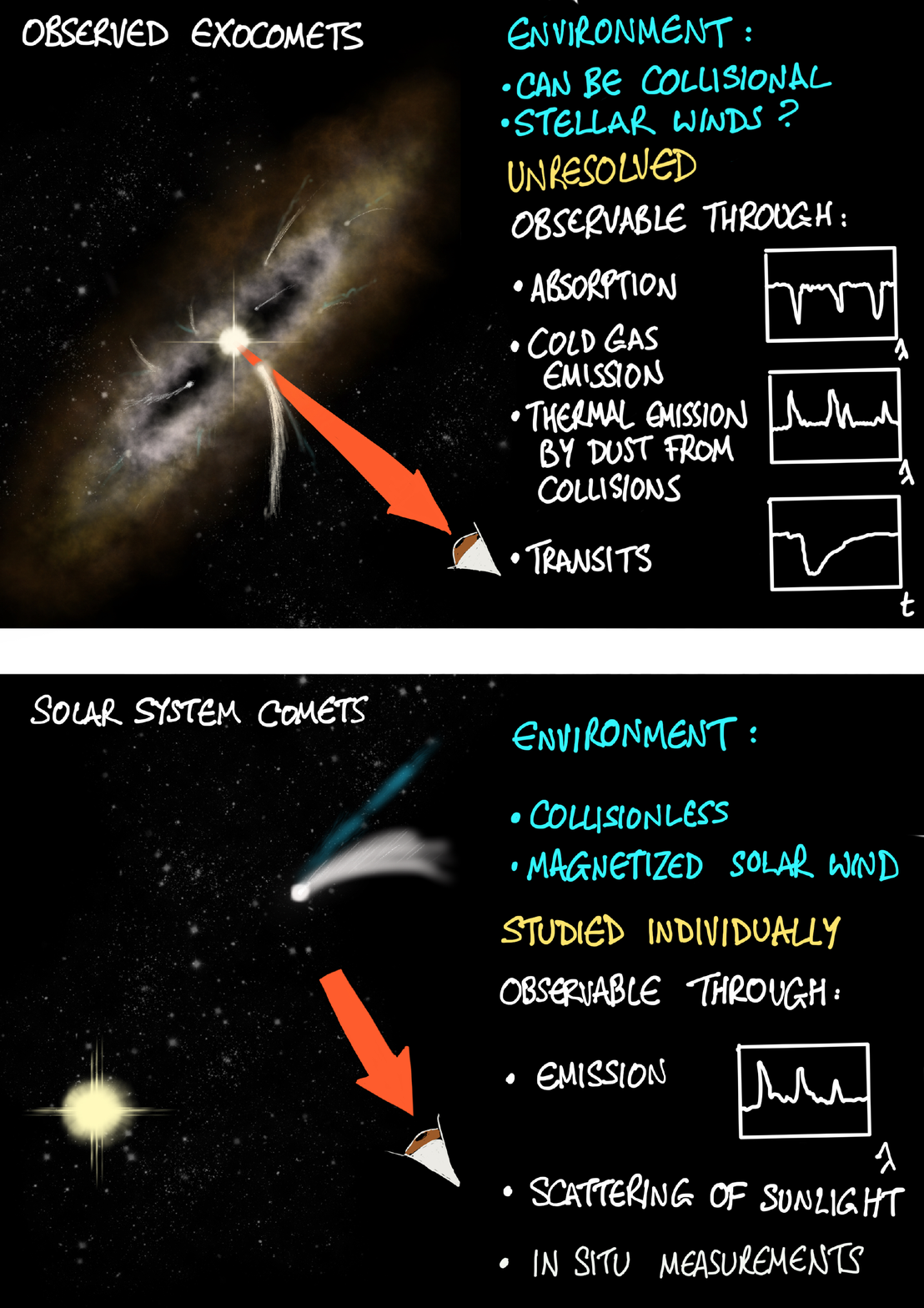}
        \caption{An overview of the environments and observational signatures of observed exocomets and Solar System comets. The differences between the collisional gas environment in which observed exocomets generally reside and the collisionless interplanetary medium in our planetary system are covered in section~\ref{sec:compositional_diff}. The solar wind is a very well-characterised medium, whereas outflowing stellar winds in observed exocomets' systems, if they exist, are very poorly characterised. The latter are extremely difficult to detect, and it is unknown whether they carry a magnetic field.}
         \label{fig:exocomet_sketch}
   \end{figure}

Comet comae in the Solar System are seen in emission. The only Solar System comet successfully observed whilst transiting the Sun's disk was the sungrazer C/2011~N3. This was only observed for a few minutes in both absorption and emission at EUV wavelengths, before the comet was apparently destroyed \citep{Schrijver2012}. To date, direct spectroscopic exocomet detections have been entirely in absorption. Lines that are commonly bright in Solar System comets (Na {\sc i} in sungrazers and a few other comets further from the Sun with very high production rates, CN in regular comets; see Table \ref{table_of_features}) have been searched for in emission in exocomets without success. Spectroscopic observations of exocomets have not yet shown the presence of CN. The exception to this is 2I/Borisov which clearly showed CN emission \citep[e.g.,][]{Fitzsimmons2019} whilst passing through our Solar System. An overview of the cometary environments and how they are observed is presented in Fig.\,\ref{fig:exocomet_sketch}.

\subsection{Compositional similarities and differences}
\label{sec:compositional_diff}

The Ca\,{\sc ii} lines commonly seen in the spectra of $\beta$~Pic and polluted WDs (see Sec.\,\ref{sec:Evidence_exocomets}) have been detected in the extreme case of the large sungrazing comet C/1965 S1 Ikeya-Seki \citep{preston1967,slaughter69}. Interestingly, the typical optical cometary emission lines (CN, C$_2$, C$_3$, etc.) were faint or undetectable in Ikeya-Seki close to its perihelion; this behaviour may provide insight into what is seen (or not seen) in exocomets close to their stars. Nitrogen-bearing molecules such as N$_2$, NH$_3$ and CN, although detectable in Solar System comets, are considered minor constituents \citep[e.g.][]{Krankowsky_1986,Eberhardt_1987,Wyckoff_1991}. Observations of the N{\sc i} line in $\beta$~Pic showed no short term variable absorption signatures of N{\sc i} which is consistent with N only being a minor exocometary consituent \citep{Wilson_2019}.

The main volatile of most Solar System comets is water ice. Observations of the Ly-$\alpha$ emission line in $\beta$~Pic showed a strong asymmetric line profile caused by additional redshifted absorption. The asymmetric line shape has been interpreted as hydrogen gas falling towards the star which may have arisen from the dissociation of water originating from sublimating exocomets \citep{Wilson_2017} or from the gas disk accreting onto the star \citep{Kral2017}.

The Na D lines at 589.0 and 589.6\,nm have been detected in over a dozen comets \citep{Cremonese2002}, primarily at heliocentric distances less than 1~au due to their very large fluorescence efficiencies for heliocentric velocities exceeding a few 10s of km/s \citep{watanabe03}. Perhaps counterintuitively, sodium has not conclusively been detected in close-in transiting exocomets. Observations of the Na D lines are challenging from the ground due to telluric contamination. Future observations to look for variable Na D lines in exocomets clearly warrants further investigation.

No subsequent sungrazing comets have approached Ikeya-Seki's size or brightness, so we have yet to study such a comet with modern instrumentation for a more direct comparison to exocomets. Comet C/2012 S1 (ISON) held great promise for such observations, as it was discovered a year before perihelion and was the subject of a worldwide observing campaign. However, it evidently began disintegrating before perihelion \citep{Knight2014}, and was undetected in the EUV around its perihelion passage \citep{Bryans2016}. More than 4000 sungrazing comets have been detected to date \citep{Battams2017}, but the vast majority are smaller than 100~m in diameter and only observed via broadband imaging \citep[e.g., ][]{Knight2010}. Limited spectroscopy of a dozen or so sungrazing comets has been obtained at UV and EUV wavelengths by solar observatories over the last two decades \citep[e.g., ][]{Bemporad2007,Schrijver2012,Bryans2012}. These observations were optimised for solar observing so the specific comet lines detected were serendipitous and not necessarily the most prominent or diagnostic. The detections of highly ionised O, C, and Fe by \emph{Solar Dynamics Observatory's} AIA instrument \citep{McCauley2013,Pesnell2014} hold promise for direct comparisons to exocomet systems since conditions are akin to star-grazing exocomet systems, but will likely require a future generation of space-based x-ray or UV facilities to be detected. The most highly ionised species attributed to exocomets are Al\,{\sc iii}, C\,{\sc iv} and Si\,{\sc iv}. Even higher ionised species may exist, but have not yet been detected. A model of cometary debris in the solar corona by \cite{Pesnell2014} opens up the possibility of detecting higher ionised species in exocomets (they model the detection of C\,{\sc iv} , Fe\,{\sc viii}  through Fe\,{\sc x}, and O\,{\sc iii} through O\,{\sc vi}. A review of observations of Near-Sun Comets in our Solar System has been provided by \citet{Jones2018}.

Solar System comet observations indicate that some intermediate charge state ions are created through photoionisation, e.g. C\,{\sc iii} detected in C/2002 X5 (Kudo-Fujikawa) when it was at $\sim$0.2~au from the Sun can be explained by the double photoionisation of C originating from cometary dust \citep{povich2003}. 
To explain the presence of highly ionised species such as Al\,{\sc iii} and C\,{\sc iv} in exocomet spectra, \citet{Beust1993} invoked the process of ionisation at a collisional shock occurring around exocomets. This is because the central star (in this case $\beta~Pic$) is unable to photoionise such highly ionised species. The heat generated by compression and collisions within the shock surface of exocomets sufficiently close to the star were thought to be high enough to explain the formation of the highly ionised ions. 

Shocks have been detected {\it in situ} at numerous Solar System comets during comet encounter missions \citep[e.g.][]{Gringauz_1986,Neubauer_1993,coates1997,Gunell_2018}. These arise due to the slow-down and deflection of the supersonic solar wind when it reaches the cometary obstacle, where significant mass, in the form of freshly-ionised cometary gas, is added to the wind. It is important to note that all solar wind shocks detected to date are collisionless rather than collisional; this is possible due to the presence of the heliospheric magnetic field that is carried by the solar wind.

Emission from higher ionisation state species (Hydrogen and Helium-like ions) has been observed in Solar System comets, too \citep{Bodewits2007}. Rather than being neutral species originating at the comet that are subsequently ionised, these ions' parent species originate at the Sun as multiply-ionised heavy ions, e.g. O$^{7,8+}$, and are carried to the comet by the solar wind \citep{Cravens1997}. Instead of moving to higher ionisation states, these are partially neutralised at the comet through charge exchange processes with neutral species in the cometary coma, as described in Section~\ref{sec:spectroscopic_features_of_comets}. 
O~{\sc iv} was detected \emph{in situ} in the ion tail of C/2006~P1 McNaught by the \emph{Ulysses} spacecraft, but it is unclear whether that ion resulted from the ionisation of cometary oxygen atoms, or the multistage neutralization of highly-charged oxygen ions in the solar wind  \citep{neugebauer07}.

Exocomets around stars with stellar winds could exhibit similar charge exchange emission at EUV and X-ray energies. This would be most apparent in charge states not expected in cold cometary environments. It is clear that stellar winds could be present at several of the systems where exocomets' presence has been inferred. Stellar winds are, however, difficult to detect; as noted by \citet{suess2015}, the Sun's solar wind would be invisible at stellar distances. The presence and nature of stellar winds in many of these systems will possibly remain undetermined. The conceivable presence of a stellar wind carrying a magnetic field, as is the case in our Solar System, should be considered when interpreting observations of the ionised components of exocomets. A decoupling in the line-of-sight velocities of neutral and ionised components of an exocomet could indicate that the ions are carried by a magnetized stellar wind, i.e., as an ion tail.

Finally, as discussed in Section \ref{sec:Exocomet_belts} and \ref{sec:Composition_Exocomets}, far-IR and millimeter observations of molecular emission lines within debris disks can be used to probe the composition of the population of exocomets within debris disks. The derived fractions of CO(+CO$_2$) ice by mass are so far found to be largely consistent with the compositions observed in Solar System comets \citep{Matra2017}, which may indicate similar formation conditions in the starting protoplanetary disks \citep{AHearn2012, Eistrup19}. In addition to the CO(+CO$_2$) ice mass fractions, mm observations have started setting upper limits on CN emission, which (assuming CN is produced by HCN photodissociation alone) set tight constraints on the HCN/(CO+CO$_2$) outgassing rate ratio. This ratio is at the low end of what is expected from typical Solar System HCN/(CO+CO$_2$) compositions in a few systems \citep{Matra2018, Kral_2020}. As mentioned in Section \ref{sec:Composition_Exocomets}, upper limits on the presence of exocometary water vapour have also been set, directly \citep{cavallius2019} and indirectly \citep{Kral2016}, within the $\beta$ Pic disk. These measurements suggest that the H$_2$O outgassing rate, when compared to that of CO(+CO$_2$), is, like HCN, also at the low end of Solar System comet range. If depletions of water and HCN are widely confirmed compared to CO(+CO$_2$), this could imply either a true depletion of H$_2$O and HCN compared to CO(+CO$_2$) ice or, given the low temperatures of tens of K at these distances, decreased outgassing for the less volatile molecules of H$_2$O and HCN compared to CO \citep{Matra2018}.

\section{Interstellar visitors}
\label{sec:interstellar_visitors}

Dynamical models suggest that a large number of our comets (as much as 90\%) was lost in the early Solar System \citep{Levison2010}, and comets are still lost through continued gravitational ejection. Similarly, other systems might eject their comets and these objects make it possible to study exocomets up close. Recently, two such interstellar objects were discovered when they passed through the Solar System, 1I/`Oumuamua and 2I/Borisov \citep[e.g.,][]{Oum2019,Fitzsimmons2019}. The discovery of two such objects suggests that future discoveries will be relatively common \citep[e.g.,][]{Trilling2017}.
 
1I/`Oumuamua passed close to the Sun (0.25\,au) and Earth (0.16\,au), and its hyperbolic orbit confirmed its interstellar origins \citep{Meech2017}. Based on its brightness, the object's effective radius was likely less than 100~m, and the large amplitude of its lightcurve implies an extreme elongated or flattened shape (reviewed in \citealt{Oum2019}). Continued observations by HST indicated that its orbit was altered by non-gravitational forces, a clear indicator for sublimation-driven activity in comets \citep{Micheli2018}. However, no evidence of a coma or tail was observed and it has been argued that the typical drivers of activity in our Solar System (H$_2$O, CO, or CO$_2$) could not have provided the observed non-gravitational acceleration for the assumed size and density \citep{Sekanina2019,seligman2020}. As a result of these unusual properties, a number of models have been suggested that are well beyond the usual paradigm for the origin of comets and asteroids in our Solar System. Several authors have suggested a combination of disruption and subsequent ejection from the host system of a large planetesimal during one or more close approaches to a giant planet, its host star, or one member of a low-mass binary \citep[e.g.,][]{Raymond2018_oumuamua,Cuk2018,Zhang2020}. Others report entirely new phenomena including an icy fractal aggregate \citep{Moro-Martin2019} or molecular hydrogen ice \citep{seligman2020}. Should future interstellar objects exhibit similar properties to `Oumuamua, it may become necessary to rethink how typical our own Solar System is.

When the second interstellar object, 2I/Borisov, was first discovered it already displayed a prominent tail, and subsequent archival searches in pre-discovery survey observations showed that it was active outside 5 au from the Sun \citep{Ye2020}. Borisov was brighter and observable for a much longer time than `Oumuamua, and the emission of several fragment species common in Solar System comets was observed, including CN, OH, C$_2$, [O~{\sc i}], and NH$_2$ \citep[e.g.,][]{Fitzsimmons2019, Xing2020, Lin_2020,McKay2020, Bannister2020}. This initially led to the conclusion that many of the properties of this object were surprisingly similar to those of comets from our Solar System. However, contemporaneous observations by HST, the \emph{Neil-Gehrels-Swift observatory}, and ALMA allowed for the measurement of the production rates of two major parent volatiles, H$_2$O and CO \citep{Bodewits2020, Cordiner2020}. They found that the object contained substantially more CO ice than H$_2$O ice, with an abundance of at least 150\%. This is very different from the CO to H$_2$O gas ratio observed in most comets in the inner Solar System, which ranges between 0.2 - 23\% and is typically around 4\% \citep{bockelee-morvan2017}. This high CO to H$_2$O ratio might be attributed to an origin in an environment significantly different from those in the early Solar System, such as around an M-dwarf star (the most abundant type of star in our galaxy, but much cooler than the Sun, see \citealt{Bodewits2020}), or from the outer regions of a protoplanetary disk, far from its host star \citep{Cordiner2020}.

The European Space Agency has selected for launch in the late 2020s the {\it Comet Interceptor} mission \citep{snodgrass2019} that could send multiple probes to an interstellar comet if a suitable target is discovered\footnote{\url{http://www.cometinterceptor.space}}. Although it is uncertain from where interstellar comets originate, they are likely to provide further insights into the similarities and differences between exocomets and Solar System comets.

\section{Summary and outlook}
\label{sec:summary_and_outlook}

In this paper we provide an overview of the observational properties of Solar System comets and exocomets and compare their similarities and differences. Compared to Solar System comets, the information we have about exocomets is very limited. While observations of exocomets are spatially unresolved and thus provide us with a holistic view, observations of Solar System comets allow us to conduct in situ observations of individual cometary components (coma, dust tail, gas tail, nucleus) at high fidelity. Despite these challenges observations of exocometary gas around main sequence stars, "polluted" WD atmospheres as well as spectroscopic observations of transiting exocomets suggest that exocomets may not be that compositionally different to Solar System comets. We assume that star and planet formation is a rather universal process, so a difference would not be that easy to explain, unless for example star/disk mass-ratios clearly differ for different targets. The detection of variable Ca\,{\sc ii} absorption lines and higher ionisation state species - in exocomets, and Solar System comets, along with the compositions found in some WD atmospheres - suggests that exocomets and comets share a similar composition. Solar system comets emit in high energy EUV and X-ray emission through the gradual neutralisation of highly charged solar wind ions. Such processes may also occur at exocomets encountering stellar winds. The presence of shocks is also detected around exocomets through observations of the variable absorption lines of highly ionised species.

Observations of interstellar visitors such as 1I/`Oumuamua and 2I/Borisov allow us to learn about the physical and chemical properties of protoplanetary disks of distant stars, although their true systems of origin are unknown to us. Compositional studies of these objects might help link the fields of exocomets and Solar System comets, and new studies of interstellar visitors hold the potential to further improve our understanding of the formation history of (exo)comets. Future observations of Na D lines in spectra of exocomet host stars will allow the similarities between exocomets and Solar System comets to be tested. If the comets are indeed similar, we expect to see variable absorption signatures in the Na D lines. Multi-wavelength photometric monitoring observations of exocometary transits will provide more information about the dust properties such as the dust reflection as a function of wavelength. Space based photometric observations with TESS and {\it Planetary Transits and Oscillations} (PLATO) are likely to provide information about the extent of the exocometary tail and will yield rough estimates of the size of the exocomets. The {\it James Webb Space Telescope} (JWST) opens up the possibility to look for new exocomet lines such as H$_2$O (6 $\mu$m), CH$_4$ (7.7 $\mu$m), C$_2$H$_2$ (13.7 $\mu$m), CO$_2$ (15 $\mu$m), and S\,{\sc i} (25.2 $\mu$m) to mention a few. Present and upcoming research facilities, both for studies in and beyond our Solar System, are expected to further bridge the cometary and exocometary science communities.

\acknowledgments

We thank the Lorentz Centre for facilitating the workshop "ExoComets: Understanding the Composition of Planetary Building Blocks" during 13-17 May 2019. The authors are grateful to the staff of the Lorentz Center for their assistance in arranging and running of the workshop, which led to the creation of this work. The authors would also like to thank Alain Lecavelier des Etangs, Grant Kennedy and Christopher Manser for feedback on the paper and for the fruitful discussions on the topic of exocomets and WDs. MK was supported by the University of Tartu ASTRA project 2014-2020.4.01.16-0029 KOMEET, financed by the EU European Regional Development Fund. FK acknowledge funding by PSL University Fellowship, and the Centre National d'Etudes Spatiales. GHJ is grateful to the UK Science and Technology Facilities Council for support through consolidated grant ST/S000240/1. DI acknowledges support from ICM (Iniciativa Cient\'ifica Milenio) via N\'ucleo Milenio de Formaci\'on Planetaria. HL acknowledges support of the Netherlands Organisation for Scientific Reseach through the Planetary and Exo-Planetary Science (PEPSci) network. LR would like to acknowledge funding from The Science and Technology Facilities Council, Jesus College (Cambridge), and The University of Cambridge. Finally we would like to thank the anonymous referee for their very thorough review of this paper.



\appendix

\section{Typical (exo)cometary spectroscopic features}
Table \ref{table_of_features} below contains some of the most common cometary emission and absorption features seen in Solar System comets. Table \ref{table_of_exocomet_features} contains a complete list of species showing variations which are attributed to the presence of exocomets.

\begin{longrotatetable}
\begin{deluxetable*}{lclll}
\tablecaption{\label{table_of_features} Typical cometary emission and absorption features throughout the electromagnetic spectrum.}
\tabletypesize{\footnotesize}
\tablehead{
\colhead{Description} &
\colhead{Wavelength} &
\colhead{Spectral lines} &
\colhead{Process} & 
\colhead{References}
}
\startdata
Gamma ray &  &  & No known emission & \\
X-ray and EUV & 100\,nm < 1 keV & Atomic ions (He\,{\sc ii}, C\,{\sc v}, C\,{\sc vi}, O\,{\sc vii}), & Solar wind charge exchange & \citet{Bodewits2007}\\
 & & atomic ions (C\,{\sc ii}, C\,{\sc iii}) & Photoionisation of atomic C & \citet{povich2003} \\
&  &  &  & \\
Far UV & 120 - 200\,nm & Atoms (H, C, O), & Fluorescence, dissociative excitation,& \citet{Feldman2018}\\
& & molecules CO, H$_2$ &electron impact excitation & \citet{Weaver:1981p12}, \cite{Combi2011}\\
&  &  &  & \\
Mid and Near UV & 200 - 380\,nm & Molecules (CO), fragments (OH, CN),& Fluorescence, dissociative excitation,& \citet{Weaver:1981p12}, \citet{Feldman2004}\\
&  & molecular ions (CO$^+$, CO$_2^+$) & electron impact excitation & \\
&  &  &  & \\
Visible & 380 - 700\,nm & Fragment species (fragments C$_2$, C$_3$, NH$_2$ and atoms, O) & Fluorescence, dissociative excitation,& \citet{Cochran2002}\\
&  &  molecular ions (H$_2$O$^+$), reflected sunlight & electron impact excitation & \citet{preston1967}, \citet{slaughter69}\\
&  & Ca\,{\sc ii} and Na\,{\sc i} (primarily for sungrazers) &  & \citet{Cremonese2002}, \citet{Douglas_1951} \\
&  &  &  & \\
Near IR & 700\,nm - 5\,$\mu$m & Dust, ice, & Reflected sunlight, Fluorescence& \citet{ootsubo2012}\\\
&  & Parent molecules (CO$_2$, H$_2$O, CO, CH$_3$OH), & electron impact excitation, & \citet{Dello_2016}\\
&  & molecular ions (CO$_2^+$), radicals (OH) & ice/mineral solid state absorption & \citet{Protopapa2014}\\
&  &  &  & \\
Mid IR & 5 - 25\,$\mu$m & Nucleus and dust & Thermal emission & \citet{fernandez2013}\\
&  &   & mineral solid-state absorption& \\
&  &  &  & \\
Far IR & 25 - 200\,$\mu$m & H$_2$O, HDO, NH$_3$, water-ice & Radiative and collisional excitation & \citet{Lellouch1998}\\
&  &   & thermal emission & \\
&  &  &  & \\
Sub-millimeter & 200\,$\mu$m - 1\,mm & Molecules (e.g. HCN, HNC, CO, CH$_3$OH, & Radiative and collisional excitation & \citet{Cordiner2014}\\
&  &  HDO, complex organics), dust & thermal emission & \citet{Har2011}\\
&  &  &  & \citet{Biver2019}\\
&  &  &  & \citet{bockelee-morvan2004}\\
&  &  &  & \\
Microwave & 1\,mm - 10\,cm & Molecules, ions, & Radiative and collisional excitation & \citet{Milam2004}\\
&  & radicals (e.g. HCN, HCO$^+$, C$_2$H, CS) &  & \\
&  &  &  & \\
Radio & $>10$\,cm & Molecules and radicals (OH, NH$_3$, H$_2$CO) & Radiative and collisional excitation & \citet{Crovisier2002}\\
&  &  &  & \citet{Howell2007}\\
\enddata
\end{deluxetable*}
\end{longrotatetable}

\begin{longrotatetable}
\begin{deluxetable*}{lclll}
\tablecaption{\label{table_of_exocomet_features} Detected exocometary absorption and emission features.}
\tablehead{
\colhead{Description} &
\colhead{Wavelength} &
\colhead{Spectral lines} &
\colhead{Process} & 
\colhead{References}
}
\startdata
Far UV & 120-200\,nm & Al\,{\sc ii}, Al\,{\sc iii}, C\,{\sc i}, C\,{\sc ii}, C\,{\sc ii}$^*$, C\,{\sc iii}, & Exocometary bow shock at a few stellar  & \cite{Deleuil1993}, \citet{Vidal-Madjar_1994}\\
&  & C\,{\sc iv}, Si\,{\sc i}, Si\,{\sc iii}, Si\,{\sc iv}, O\,{\sc i} & radii and sublimation of dust grains & \citet{Miles2016}, \citet{Grady2018}\\
&  & Ni\,{\sc ii} ,S\,{\sc i}, N\,{\sc i}, CO  & Molecules and photodissociation products & \citet{Lagrange1998}, \citet{Wilson_2019} \\
&  &  & from release in exocometary belts  & \citet{Roberge2000}\\
&  &  &  & \\
Mid and Near UV & 200-400\,nm & Mg\,{\sc ii}, Fe\,{\sc i}, Fe\,{\sc ii}, Cr\,{\sc ii}, Mn\,{\sc ii}, Zn\,{\sc ii}, & Sublimation of dust grains at several & \citet{Vidal-Madjar_1994}, \citet{Kiefer_2019}\\
&  &  & tens of stellar radii & \citet{Lagrange1998} \\
&  &  &  & \\
Visible & 380-700\,nm & Ca\,{\sc ii}, Na\,{\sc i}, & Sublimation of dust grains at several & \citet{Ferlet_1987}, \citet{Brandeker2004}\\
&  &  & tens of stellar radii. Gas release in & \\
&  &  & exocometary belts (Na\,{\sc i}) & \\
&  &  &  & \\
Far IR & 25-200\,$\mu$m & O\,{\sc i}, C\,{\sc ii} & Photodissociation of CO in exocometary belts& \citet{Cataldi2014}, \citet{Kral2016}\\
&  &  & forming neutral oxygen and ionised carbon & \\
&  &  &  & \\
Sub-millimeter & 500\,$\mu$m - 1.3\,mm & C{\sc i}, CO & Cold CO produced in collisionally active exocometary & \citet{Moor2017}, \citet{Higuchi2017}\\
&  &  &   belts and its photodissociation product neutral carbon & \citet{Kral2019} \\
&  &  &  & \\
\enddata
\end{deluxetable*}
\end{longrotatetable}

\end{CJK}

\bibliography{references}{}
\bibliographystyle{aasjournal}



\end{document}